\documentclass[aps,prb,reprint,longbibliography]{revtex4-2}

\usepackage{graphicx}   
\usepackage{dcolumn}    
\usepackage{bm}         
\usepackage{amsmath}    
\usepackage{amssymb}    
\usepackage{mathtools}

\newcommand{\bracket}[1]{{\left< #1 \right>}}

\begin{document}

\title{Moiré magnetism in a bilayer Ising model}

\author{Ryan Flynn}
\email{rflynn22@bu.edu}
\affiliation{Department of Physics, Boston University, 590 Commonwealth Avenue, Boston, Massachusetts 02215, USA}

\author{Anders W. Sandvik}
\email{sandvik@bu.edu}
\affiliation{Department of Physics, Boston University, 590 Commonwealth Avenue, Boston, Massachusetts 02215, USA}

\date{\today}

\begin{abstract}
Moiré patterns in magnetic bilayers generate spatially modulated interlayer exchange interactions that can give rise to nonuniform magnetic textures. We study a minimal classical bilayer Ising model with a moiré-modulated interlayer coupling, generated either by relative twist or differential strain between the layers. Using large-scale classical Monte Carlo simulations, we show that the ordering transition remains in the conventional two-dimensional Ising universality class, even when the low-temperature state is domain-textured. At low temperatures, we find a smooth crossover between a uniform ferromagnet and domain-textured state, in which the spins locally follow the sign of the interlayer exchange. We demonstrate that there is no breaking of layer symmetry for twisted bilayers. The location of the crossover is determined by a simple geometric energy balance between bulk interlayer exchange and intralayer domain-wall costs. Our results provide a minimal framework for understanding how moiré-modulated magnetic textures can emerge from geometric energetics without requiring a thermodynamic phase transition.
\end{abstract}

\maketitle

\section{Introduction}

Moiré structures in stacks of two-dimensional materials have emerged as a versatile platform for engineering their electronic and magnetic properties~\cite{Novoselov2016,Gibertini2019}. This has been particularly true in electronic systems such as twisted bilayer graphene, where superconductivity has been observed as the moiré pattern is tuned~\cite{Bistritzer2011,Cao2018a}. In magnetic bilayers, the same geometric mechanisms - relative twist, lattice mismatch, or strain between layers---create a moiré pattern that leads to spatially modulated interlayer exchange interactions \cite{Hejazi2020,Sivadas2018}. This moiré pattern has been shown to give rise to a variety of spatially nonuniform magnetic textures in both theory and experiment, including noncollinear and domain-like states in twisted bilayer CrI$_3$ and related materials \cite{Kim2022,Cheng2023,Xu2022}.

Two-dimensional moiré ferromagnetism was first experimentally demonstrated in monolayer crystals of CrI$_3$ and Cr$_2$Ge$_2$Te$_6$ \cite{McGuire2014,Samarth2017,Gong2017}. This realization, among other applications, has enabled the exploration of magnetic effects due to moiré patterns in heterostructures of these materials \cite{Burch2018,Park2016}. Early work demonstrated that magnetic properties in layered magnets such as CrI$_3$ can be tuned by external electric fields, establishing interlayer exchange as a sensitive and controllable degree of freedom \cite{Huang2018,Cheng2022}.

In a continuum field theory model, Hejazi \textit{et al.} \cite{Hejazi2020} demonstrated that both relative twist and strain between layers produce a spatially varying interlayer exchange with the periodicity of the moiré superlattice in stacked ferromagnets and antiferromagnets \cite{Hejazi2020,Balents2019}. This moiré modulation was shown to give rise to nonuniform magnetic phases, including noncollinear states. Low temperature transitions between these phases can be driven by tuning the twist angle or strain. These predictions have since been supported by experimental observations in twisted magnetic bilayers \cite{Cheng2023, Jiang2019, Song2021, Xu2022}.  First-principles calculations for bilayer CrI$_3$ have further established the dependence of interlayer exchange on stacking order, providing a microscopic basis for moiré-modulated magnetism in real materials \cite{Sivadas2018,Soriano2019}.

While twisting layers has been the dominant experimental route to realizing magnetic moiré structures, similar modulation can also be engineered solely through differential strain between the layers. This has been proposed recently both theoretically~\cite{Escudero2024} and experimentally~\cite{Yao2024}, and provides another potentially controllable mechanism to tune the moiré pattern. This method mimics the intrinsic moiré patterns created by stacking materials with naturally different lattice constants, such as in the case of graphene on hexagonal boron nitride~\cite{Jung2014}.

In many moiré magnetic systems, spatially modulated magnetic textures emerge as system parameters such as twist angle or strain are varied. However, the appearance of such textures does not necessarily imply the existence of a distinct thermodynamic phase. Instead, changes in the characteristic spatial structure can occur as smooth crossovers without any additional spontaneous symmetry breaking. 

In this work, we study a minimal classical bilayer Ising model with a moiré-modulated interlayer exchange, which isolates the essential geometric ingredient of moiré magnetism without aiming to model any specific material. The simplicity allows for easy simulation via classical Monte Carlo methods, and allows us to study the low temperature magnetic textures that emerge. 

We show that at low temperatures the system supports two distinct magnetic states: a uniform ferromagnet and a domain-textured state in which spins locally follow the sign of the interlayer exchange. Importantly, these configurations belong to the same thermodynamic phase, sharing the same broken $\mathbb{Z}_2$ symmetry, and differ only in their spatial structure. We show that no additional symmetry is broken by the domain-textured state, and thus find no evidence of a first- or second-order transition associated with this crossover. 

Finally, we show that the location of the crossover is governed by a simple geometric energy balance between bulk interlayer exchange and intralayer domain-wall costs, leading to a predictable dependence on the moiré parameter and interlayer coupling strength. This yields an energetic crossover map that applies equally to strained and twisted bilayers, with the only qualitative distinction arising from explicit layer-symmetry breaking in the strained case. 

\begin{figure}[t]
    \centering
    \includegraphics[width = \columnwidth]{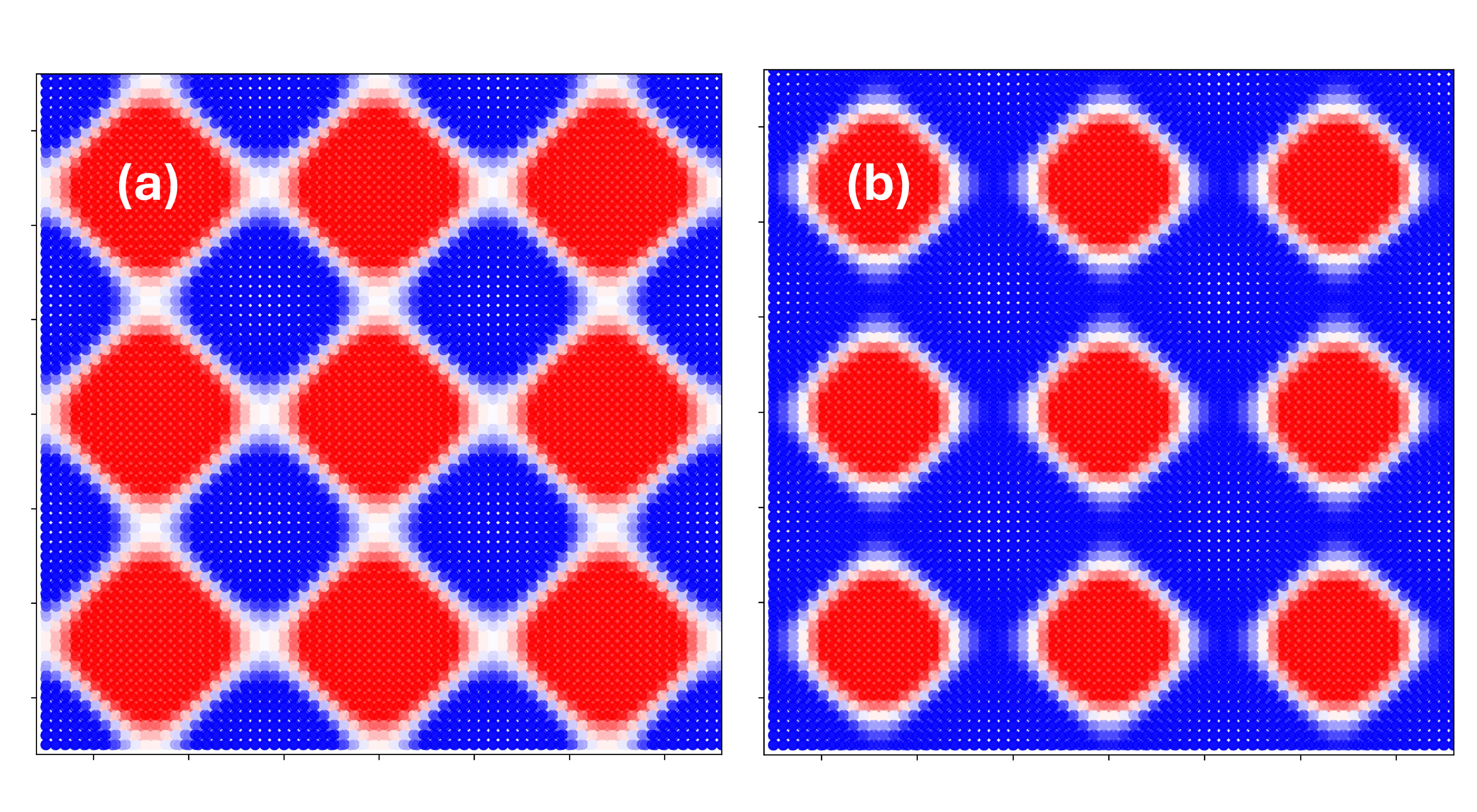}
    \caption{Map of the interlayer coupling function $\Phi(\mathbf{u}) = \Phi_0 + \cos(\mathbf{b}_1\cdot\mathbf{u}) + \cos(\mathbf{b}_2\cdot\mathbf{u})$, with $\Phi_0 = 0$ in (a), and $\Phi_0 = 0.5$ in (b), and where $\mathbf{b}_a$ are the reciprocal lattice vectors and $\mathbf{u}$ is the displacement field.  The lattice is size $L=72$ with differential strain $a = 24/23$, yielding $N_M=3$ (linear) moiré unit cells. Blue (red) represents where the interlayer exchange is (anti)ferromagnetic. The constant $\Phi_0$ in the interlayer coupling function ``softens'' the domains. The overall coupling strength is controlled by the parameter $J'$.}
    \label{fig:moire}
\end{figure}

\section{Methods}

Our system is a square-lattice bilayer with Ising spins at each site. A moiré pattern is generated by deforming one layer relative to the other, either through a relative twist or differential strain. Both deformations produce qualitatively similar long-wavelength moiré patterns; however, differential strain explicitly breaks the symmetry of exchanging the two layers, while a twisted bilayer preserves it. Twisted bilayers have been the dominant experimental realization of moiré magnetism, while differential strain remains comparatively less explored. In this work, we consider both methods. 

We formulate the model on a square lattice for simplicity, as a minimal model that isolates the geometric mechanism of moiré modulation from material-specific details. Our main conclusions---a continuous phase transition demonstrated by a test of Ising universality and the crossover nature of domain formation---depend on the competition between intralayer and interlayer exchange and are expected to hold for other lattice geometries, including the honeycomb lattice relevant to materials such as CrI$_3$.

For a twisted bilayer, the control or moiré parameter is the twist angle $\phi$. For differential strain, we define the parameter $a$ as the ratio of the lattice constant of one layer relative to the other. Without loss of generality, we will always take $a\geq1$, with $a=1$ corresponding to undeformed layers and the absence of a moiré pattern. The size of a moiré unit cell $L_M$ in terms of the control parameters is
\begin{equation}\label{eq:unitcell}
    L_M = \frac{a}{a-1} = \frac{1}{\tan{\phi}},
\end{equation}
which also gives us a relationship between the two parameters. As $a\rightarrow1$ or $\phi\rightarrow0$, the moiré unit cell size grows.

The model Hamiltonian is given by
\begin{eqnarray}\label{eq:Ham}
  H =~ & & -J\sum_{\bracket{ij},l}\sigma_{i,l}\sigma_{j,l} \\
      & & - J'\sum_{\mathcal{N}\left[ij\right]}e^{-|\mathbf{r}_{ij}|/r_c}\Phi(\mathbf{u}(\mathbf{r}_{ij}))\sigma_{i,1}\sigma_{j,2}, \nonumber
\end{eqnarray}
where first term is a ferromagnetic nearest-neighbor ($\bracket{ij}$) Ising interaction, acting only between spins in the same layer. The strength of this interaction is set to $J=1$, with energies measured in units of $J$. The interlayer coupling strength $J'$ is an additional control parameter. $\mathcal{N}\left[ij\right]$ denotes that spins on sites $i,j$ are in the same ``neighborhood'', defined such that the projected distance between the two spins $|\mathbf{r}_{ij}| \equiv |\mathbf{r}_i-\mathbf{r}_j| \le r_c$ for $\mathbf{r}_i$, $\mathbf{r}_j$ the coordinates of the spins. $r_c$ controls the interlayer connectivity and range of the interaction, and in this work we set $r_c = \sqrt{2}$ to keep this exchange local. The interlayer coupling function $\Phi(\mathbf{u})$ is
\begin{equation}
    \Phi(\mathbf{u}) = \Phi_0 + \cos(\mathbf{b}_1\cdot\mathbf{u}) + \cos(\mathbf{b}_2\cdot\mathbf{u}),
\end{equation}
which is motivated by the form for a ferromagnetic bilayer from Hejazi \textit{et al.} ~\cite{Hejazi2020}. In the coupling function, the vectors $\mathbf{b}_1 = \frac{2\pi}{a}\mathbf{x}$ and $\mathbf{b}_2 = \frac{2\pi}{a}\mathbf{y}$ are the lowest reciprocal lattice harmonics, and $\mathbf{u}$ is the displacement field between the two layers, which encodes the local stacking offset between layers. Its form depends on the method of deformation, as described below.

The effect of this coupling function is that the interlayer interaction varies spatially with the periodicity of the moiré superlattice, alternating between ferro- and antiferromagnetic exchange. Higher harmonics are neglected as they do not qualitatively affect the resulting magnetic textures. The constant term $\Phi_0$  controls the net ferromagnetic bias between layers, and we treat this constant as a free parameter. To model a specific material, $\Phi_0$ would be set using a first-principles calculation. 

The coupling function and moiré pattern cause frustration in the system, with competition between the intralayer and interlayer exchange. We can tune which interaction dominates using the control parameters $J'$ and $a/\phi$. In Fig.~\ref{fig:moire}, we show a map of the coupling function for a strained bilayer with two different values of the constant $\Phi_0 = 0, 1/2$ and $a = 24/23$. $\Phi_0$ controls the size of the antiferromagnetic regions but does not otherwise change the moiré length scale. 

The moiré modulation is implemented differently for the strained and twisted bilayers. For the strained bilayer, the two layers are constructed with different lattice constants. The bottom layer has unit lattice spacing, while the top layer has lattice constant $a > 1$, resulting in a reduced number of spins in the strained layer in a finite sized system. Note that this is biaxial strain and preserves the square-lattice rotational symmetry. As a consequence, spins in different layers are no longer in one-to-one correspondence. This construction produces a perfectly periodic moiré pattern while explicitly breaking the layer symmetry, which correctly models differential strain. The displacement vector is then $\mathbf{u} = \mathbf{r}_i-\mathbf{r}_j$, and we can easily define the full interlayer coupling from Eq.~(\ref{eq:Ham}) between all interlayer pairs satisfying $|\mathbf{r}_{ij}| \le r_c$, or $\mathcal{N}\left[ij\right]$. The exponential factor in Eq.~(\ref{eq:Ham}) further screens the interaction, ensuring the interlayer exchange remains local within the cutoff radius $r_c$.

\begin{figure}[t]
    \centering
    \includegraphics[width = 0.95\columnwidth]{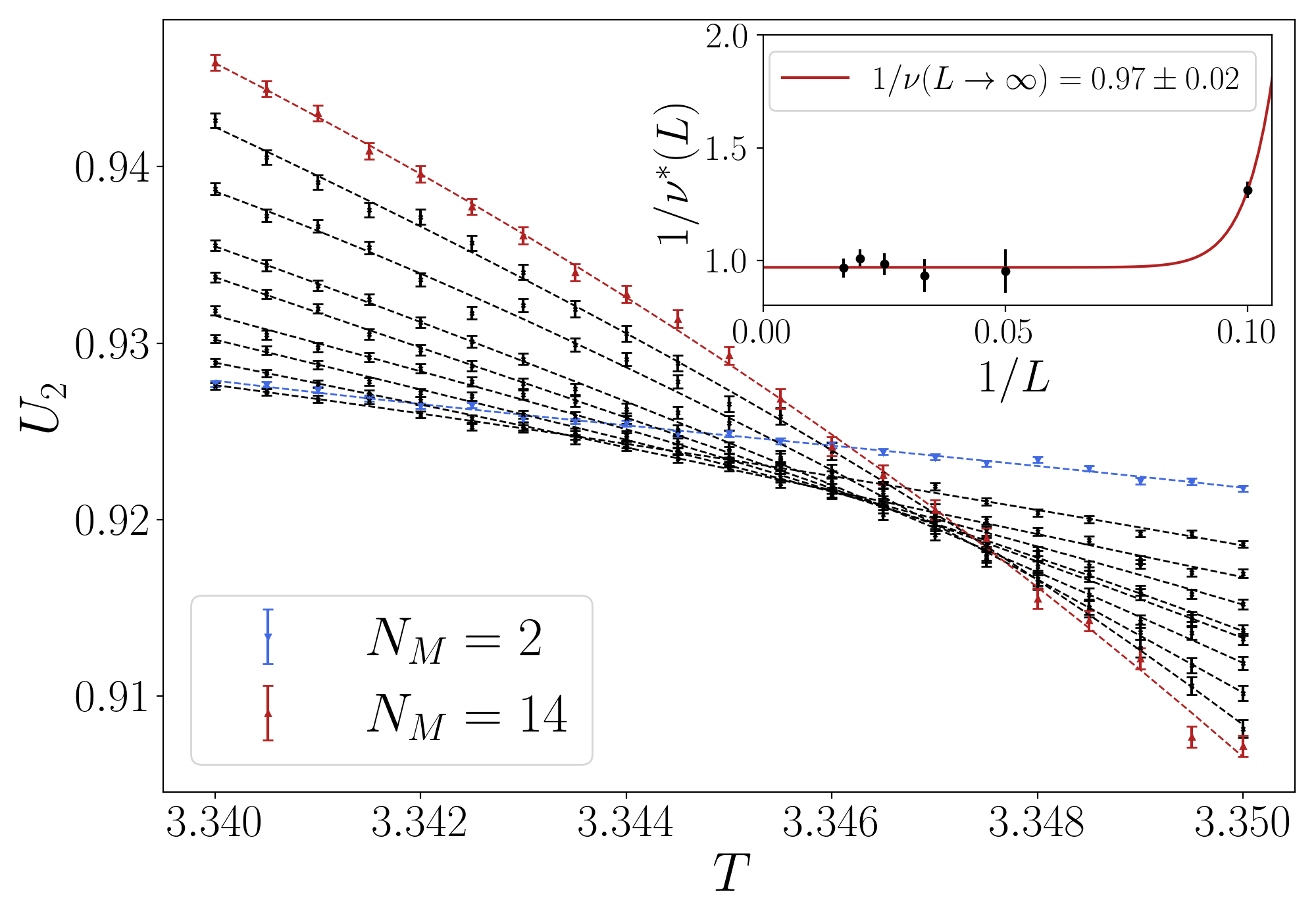}
    \caption{Binder cumulant crossings for the temperature-driven transition from the paramagnetic phase to the ordered, domain-textured state in a strained bilayer. The moiré unit cell size is $L_M = 10$ sites, corresponding to $a = 10/9$, and $N_M=L/L_M$ denotes the number of moiré unit cells along each linear system dimension. Cumulants shown for $N_M = 2,\ldots,14$ in increments of $1$. The bias $\Phi_0 = 0.5$. Lines are cubic fits to the Binder cumulant data and are shown as guides to the eye. (Inset) The fitted slopes of $U_2$ at the crossing point (see Sec.~\ref{sec:Temp}) are used to compute the size-dependent effective exponent $1/\nu^*(L)$, which extrapolates to $1/\nu = 0.97 \pm 0.02$ in the thermodynamic limit, consistent with the conventional 2D Ising universality class.}
    \label{fig:binderdm}
\end{figure}

For the twisted bilayer, the two layers are kept on identical square lattices with equal lattice constants. Explicitly rotating one or both layers relative to the other would yield non-periodic finite-size lattices. Instead, we compute the moiré-modulated coupling function corresponding to a twist angle $\phi$. In this case, the displacement vector is $\mathbf{u} = \frac{\phi}{2}\mathbf{\hat{z}}\times\mathbf{r}$ \cite{Hejazi2020}. We then overlay this coupling map onto the regular bilayer lattice. In this approach, the spins themselves are not displaced as above, but the spatially varying interlayer exchange encodes the effect of twisting. This procedure preserves periodic boundary conditions, which would otherwise be lost, and the exact layer symmetry. In this case, spins between layers \emph{are} in one-to-one correspondence, and so the cutoff radius and exponential screen are unnecessary for the chosen $r_c$. As the interlayer coupling function is what encodes the moiré physics, this procedure does not fundamentally alter the results and allows for finite-size scaling analysis. 

In van der Waals materials, the interlayer coupling strength is typically much weaker than that within layers, $J' \ll J$. In order to work with reasonably sized lattices and observe the crossover to the domain state, we consider values $J' = \mathcal{O}(J)$. We show in Sec.~\ref{sec:Cross} that this is valid as we can extrapolate along the crossover boundary to smaller $J'$ and larger moiré unit cells. 

We study the model using classical Monte Carlo simulations with both Metropolis single-spin and Swendsen Wang cluster updates. In the low-temperature ordered regime, equilibration is achieved by local rearrangements of domain walls and by tunneling domains between layers. These processes are not efficiently sampled by cluster algorithms at low-temperatures, where typical cluster sizes span the entire system. Single spin-flip updates allow these local processes and are thus better suited for equilibrating the domain state. For the ordering transition at higher temperatures, cluster updates are effective despite the frustrated interactions. Simulations are performed on square lattices of size $L$ with periodic boundary conditions in both spatial directions for both layers. 

We vary the temperature, interlayer coupling strength $J'$, and deformation parameters $a$ and $\phi$ to study both the thermal ordering transition and the low-temperature crossover from the ferromagnetic to domain state. At low temperatures, particularly in the domain-textured regime, the system exhibits long-lived metastable configurations due to the energy cost of rearranging domain walls. 

\begin{figure}[t]
    \centering
    \includegraphics[width=0.95\columnwidth]{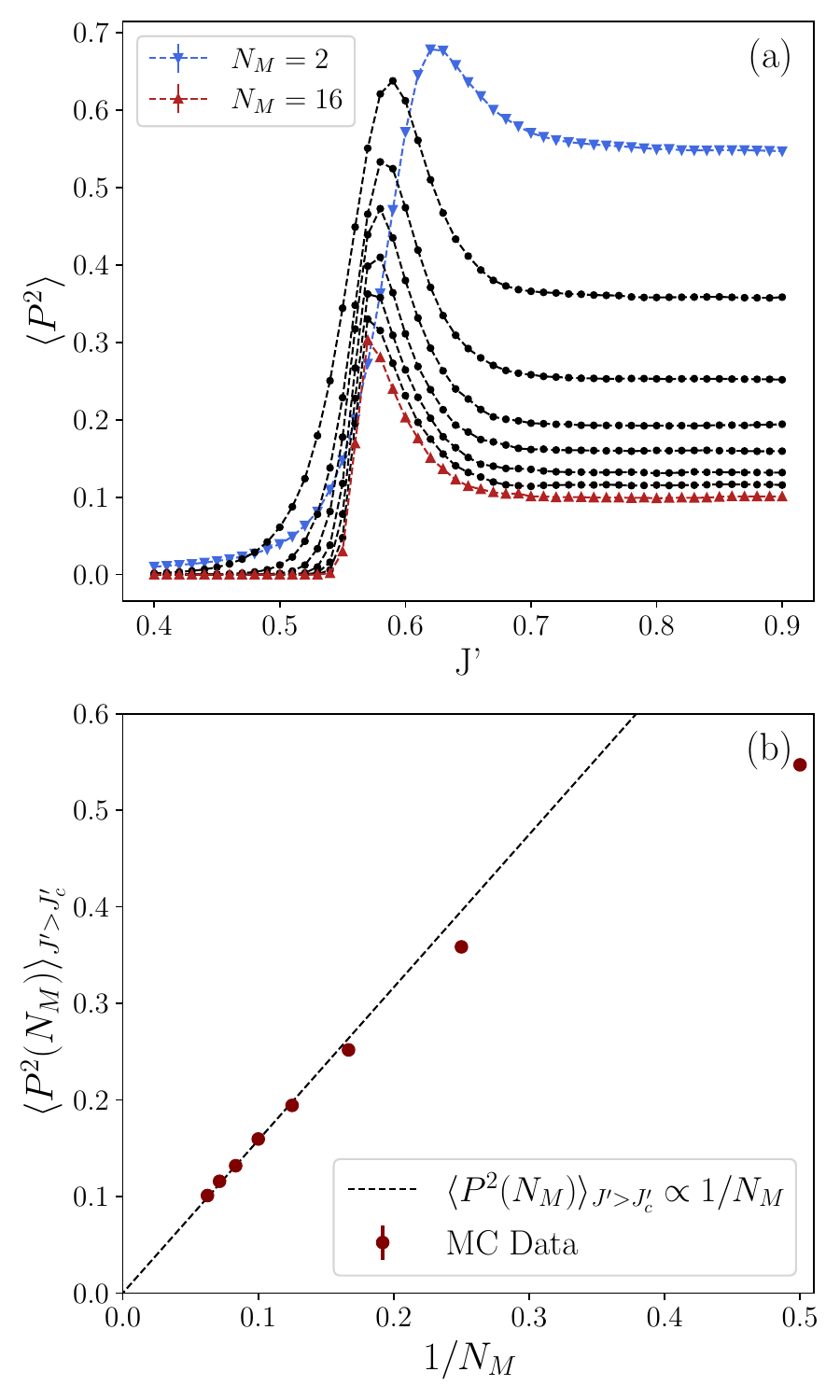}
    \caption{No layer symmetry breaking in the twisted bilayer, with twist angle $\phi = 0.1$ ($L_M = 10$) and bias $\Phi_0 = 0.2$. (a) Layer polarization order parameter $\langle P^2\rangle$ versus $J'$ for several system sizes, labeled by the number of moiré unit cells $N_M = L/L_M$ ($N_M = 2,\ldots,16$, only even). $\langle P^2\rangle$ peaks near the crossover $J'_c$, and approaches a size-dependent plateau at large $J'$. (b) The plateau values $\langle P^2(N_M)\rangle_{J'>J'_c}$ decrease with system size and extrapolate to zero in the thermodynamic limit. The largest system sizes ($N_M > 8$) follow the expected scaling $\langle P^2\rangle\propto 1/N_M$.}
    \label{fig:layersymm}
\end{figure}

\section{Temperature-Driven Transition}\label{sec:Temp}

We first analyze the ordering transition in the regime where the low-temperature state exhibits a moiré-induced domain texture. In the presence of spatially modulated and competing interactions, it is not a priori guaranteed that the ordering transition remains identical to that of the uniform Ising model, as there are other symmetries that may be simultaneously and spontaneously broken. Before turning to the evolution of magnetic textures within the ordered phase, we establish the thermodynamic phase structure of the model.

To this end, we analyze the Binder cumulant,
\begin{equation}
    U_2 = \frac{3}{2}\left(1 - \frac{1}{3}\frac{\langle m^4\rangle}{\langle m^2 \rangle^2}\right),
\end{equation}
where $m$ is the total magnetization summed over both layers. The constant $3/2$ normalizes $U_2$ to range from $0$ (disorder) to $1$ (order). The Binder cumulant probes the breaking of the global spin-flip $\mathbb{Z}_2$ symmetry.

Figure~\ref{fig:binderdm} shows the Binder cumulant as a function of temperature for several system sizes, expressed in terms of the number of moiré unit cells $N_M = L/L_M$, for a strained bilayer with large $J'$ such that domains form in the ordered phase. The ferromagnetic bias is set to $\Phi_0 = 0.5$, and this value only changes the non-universal value of the critical temperature $T_c$.  For these parameters, we observe a single, well-defined crossing at $T_c = 3.347$, indicating a continuous transition from the paramagnetic phase directly into an ordered state. This behavior arises although the low-temperature configuration is strongly domain-textured, demonstrating that the emergence of moiré-induced domains is not associated with an additional thermodynamic phase transition.

To further characterize the transition, we extract the correlation length exponent $\nu$ from the finite-size scaling of the Binder cumulant slope. Near criticality, the derivative obeys
\begin{equation}
    \frac{dU_2}{dT} \propto L^{1/\nu}.
\end{equation}
Using size-pairs ($L$,$2L$), we can obtain an effective exponent $1/\nu^*(L)$ that converges with increasing system size, which we show in the inset of Fig.~\ref{fig:binderdm}. The full derivation of this method can be found in the appendix of Shao et al.~\cite{Shao2016}. Extrapolation to the thermodynamic limit yields $1/\nu = 0.97 \pm 0.02$ in excellent agreement with the 2D Ising universality class value. 

These results demonstrate that, despite the presence of long-wavelength moiré modulation and competing ferro- and antiferromagnetic interactions, the thermal ordering transition remains in the Ising universality class. While the preserved universality class may not be surprising, the transition could in principle have been first-order, which the expected universal Ising scaling behavior rules out. The formation of spatially modulated domain textures therefore occurs entirely within the ordered phase, motivating the analysis of this crossover behavior below $T_c$.

\section{Low-Temperature Crossover}\label{sec:Cross}

\begin{figure*}
    \centering
    \includegraphics[width=\textwidth]{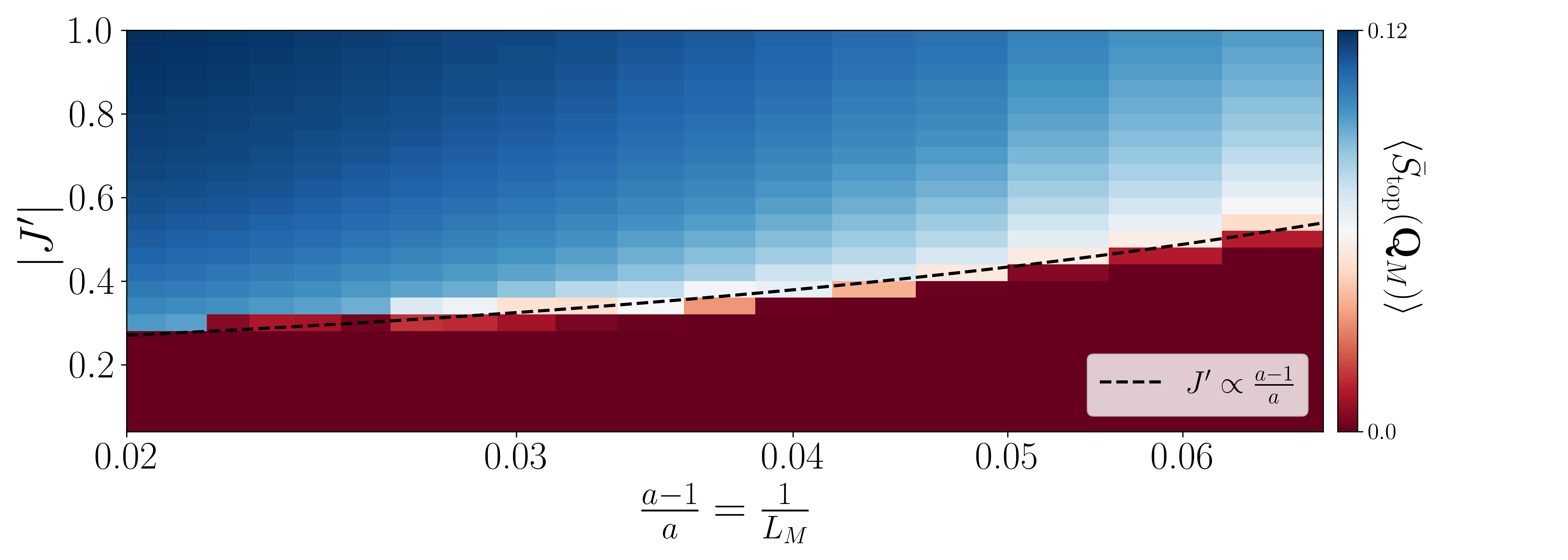}
    \caption{Crossover diagram in the interlayer coupling $J'$ and differential strain $a$, which we express as the ratio $(a-1)/a = 1/L_M$, showing the crossover from the ferromagnetic to domain-textured state. The bias is $\Phi_0 = 0.5$. This occurs at low temperature; here $T=1\ll T_c$. The color denotes the structure factor averaged over all moiré wave vectors, defined in Eq.~(\ref{Eq:StructureFactor}). The crossover boundary is fit with the linear function $J' \propto (a-1)/a$ (dashed line), which emerges from geometric energy balance of the domains.}
    \label{fig:crossover}
\end{figure*}

Having established that the system undergoes a conventional Ising transition into the ordered phase for all values of the moiré parameters, we now turn to the evolution of domains at low temperatures. The competition between the intralayer ferromagnetic exchange and moiré-modulated interlayer coupling gives rise to two distinct states below $T_c$: a uniform ferromagnet and a domain-textured state in which the spins locally align with the coupling function $\Phi(\mathbf{u})$. In this domain state, domain walls form in each moiré unit cell in one of the two layers. We show here that these states are not separated by any thermodynamic phase transition, and instead there is a smooth crossover governed by the energy balance between the domain walls and bulk. 

In the ordered phase $T<T_c$, there are few remaining symmetries that could be broken to yield a true phase transition from the ferromagnetic to domain state. Lattice translational symmetry is explicitly reduced to the moiré superlattice scale by the coupling function $\Phi(\mathbf{u})$, which depends implicitly on the moiré wavevector $Q_M$. We therefore focus on the remaining layer-exchange symmetry $\{\sigma\}_{l=1}\leftrightarrow\{\sigma\}_{l=2}$. In the strained bilayer, this symmetry is explicitly broken due to the unequal spin density in the layers, and all domain walls form in the layer with larger lattice constant (the top layer in our simulations). However, for the twisted bilayer the Hamiltonian is layer-symmetric and conceivably an effective interaction between domains could spontaneously break this layer symmetry and generate a new phase. We therefore focus here on the twisted bilayer. 

First, to distinguish between the ferromagnetic and domain states we will use the static structure factor for a layer $l$
\begin{equation}\label{Eq:StructureFactor}
    S_l(\mathbf{q}) = \frac{1}{N_l^2}\left|\sum_{i\in l}\sigma_i e^{i\mathbf{q\cdot r}}\right|^2
\end{equation}
where $N_l$ is the number of spins per layer. We can evaluate the static structure factor at the moiré reciprocal lattice vectors, $\mathbf{Q}_M = (\pm 2\pi/L_M,0), (0,\pm 2\pi/L_M)$, to obtain $\bar{S}_l(\mathbf{Q}_M)$, where the bar indicates averaging over the four $\mathbf{Q}_M$. $\bar{S}_l(\mathbf{Q}_M)$ will be zero in the ferromagnetic state and nonzero in the domain state, and thus serves as an ``order parameter'' to study the crossover. 

We can further define a polarization order parameter
\begin{equation}
    P = \frac{\bar{S_2}(\mathbf{Q}_M) - \bar{S_1}(\mathbf{Q}_M)}{\bar{S_2}(\mathbf{Q}_M)+\bar{S_1}(\mathbf{Q}_M)}.
\end{equation}
Where the subscripts $1$ and $2$ indicate bottom and top layers respectively. This polarization parameter can be used to measure the presence of any layer-symmetry breaking. In Fig.~\ref{fig:layersymm}, we show $\langle P^2\rangle$ as $J'$ is tuned and domain walls form in the system at low temperature $T=1 \ll T_c$, well below the ordering transition. We label system sizes by the number of moiré unit cells $N_M$. For no layer-symmetry breaking, we expect $\langle P^2 \rangle$ to be zero both above and below the crossover point $J'_c$ in the limit $N_M\rightarrow\infty$. The constant plateau in the domain state above the crossover $\langle P^2(N_M)\rangle_{J'>J'_c}$ is due to finite-size fluctuations. For no layer-symmetry breaking, the layer polarization $\langle P\rangle$ should be binomially distributed, as each domain independently forms in the top or bottom layer. In this case, we expect $\langle P^2(N_M)\rangle_{J'>J'_c}$ to scale inversely with the number of moiré unit cells as $1/N_M$. We show that this is the case in the lower panel of Fig.~\ref{fig:layersymm},  and therefore in the thermodynamic limit $\langle P^2(L\rightarrow\infty)\rangle_{J'>J'_c}\rightarrow0$. We conclude that the twisted bilayer does not break the layer symmetry. Therefore, the strained bilayer (in which layer symmetry is explicitly broken) and the twisted bilayer (in which it remains unbroken) both exhibit a smooth crossover and a single low-temperature phase. 

In Fig.~\ref{fig:crossover} we show the crossover boundary between the two low-temperature states in the strained bilayer. To detect onset of the moiré domain texture we wish to detect the translational symmetry breaking. Since we consider the strained bilayer, all domains will form in the top layer, and we measure $\bar{S}_\text{top}(\mathbf{Q}_M)$ as defined in Eq. \ref{Eq:StructureFactor}. As demonstrated before, the twisted case will also exhibit a crossover, and we instead measure $\bar{S}_{\text{tot}}(\mathbf{Q}_M)$. We can explain the boundary with a simple energetic argument. The bulk interlayer energy of the domains scales as $J' L_M^2$, which we can relate to the moiré parameters with Eq.~(\ref{eq:unitcell}). The domain-wall energy scales as $JL_M$ (with $J\equiv 1$). The crossover boundary should then follow a simple relationship 
\begin{equation}
    J' \propto 1/L_M.
\end{equation}
The constant of proportionality will depend on system parameters, such as the ferromagnetic bias $\Phi_0$ and temperature. For example, increasing $\Phi_0$ decreases the effective size of the domains (see Fig.~\ref{fig:moire}) and thus would shift the boundary but maintain the simple form. Given Eq.~(\ref{eq:unitcell}), we fit the boundary function $J'\propto (a-1)/a = 1/L_M$ in Fig.~\ref{fig:crossover} and find it is consistent with our numerical results. For the twisted bilayer, this maps to $J'\propto \tan\phi$. 

We therefore justify our initial choice $J'=\mathcal{O}(J)$, as we show that these results can be extrapolated down to the regime of small interlayer coupling, $J'\ll J$, provided the moiré unit cell size is large, $L_M \gg 1$.

\section{Conclusion}

We now provide a summary of our model and numerical results. 

\begin{itemize}
    \item We introduce a minimal Ising model to study moiré magnetism in a square lattice bilayer. We generate the moiré pattern via either twist or differential strain, and find they create qualitatively similar interlayer exchange couplings. 

    \item We show that the temperature-driven phase transition from the paramagnetic to ordered phase is a conventional Ising transition, regardless of whether the low-temperature state is ferromagnetic or domain-textured. 
    
    \item We determine that for both twisted and strained bilayers there is no thermodynamic phase transition, but rather a smooth crossover, between the ferromagnet and domain state at low temperatures. The crossover boundary is characterized solely by the energetic balance between domain walls and bulk. In the strained case, this is due to the layer symmetry being explicitly broken, while for the twisted bilayer it remains intact through the crossover. 
\end{itemize}

By focusing on a minimal Ising model, we emphasize the geometric effects of both twist and strain in low-temperature magnetically ordered phases. In both cases, the appearance of ferro- and antiferromagnetic regions occurs, which is qualitatively consistent with experimental observations of materials such as CrI$_3$. Our results clarify that the appearance of moiré-induced magnetic textures does not by itself imply the existence of a distinct thermodynamic phase.

\section{Acknowledgments}

We would like to thank Guanghui Cheng and Gabe Schumm for stimulating discussions. This research was supported
by the Simons Foundation under Grant No. 511064. The
numerical calculations were carried out on the Shared
Computing Cluster managed by Boston University’s Research Computing Services.

\vfill

\bibliography{bib.bib}

\end{document}